\documentclass[sigconf, authorversion, nonacm]{acmart}
\AtBeginDocument{%
  \providecommand\BibTeX{{%
    \normalfont B\kern-0.5em{\scshape i\kern-0.25em b}\kern-0.8em\TeX}}}
\acmConference[SPAA '24]{June 17-21,
  2024}{Nantes, France}

\begin{document}

\title{Efficient Parallel Reinforcement Learning Framework using the Reactor Model}

\author{Jacky Kwok}
\affiliation{%
  \institution{UC Berkeley}
  \city{Berkeley}
  \country{United States}}
\email{jackykwok@berkeley.edu}

\author{Marten Lohstroh}
\affiliation{%
  \institution{UC Berkeley}
  \city{Berkeley}
  \country{United States}}
\email{marten@berkeley.edu}

\author{Edward A. Lee}
\affiliation{%
  \institution{UC Berkeley}
  \city{Berkeley}
  \country{United States}}
\email{eal@berkeley.edu}


\begin{abstract}
Parallel Reinforcement Learning (RL) frameworks are essential for mapping RL workloads to multiple computational resources, allowing for faster generation of samples, estimation of values, and policy improvement. These computational paradigms require a seamless integration of training, serving, and simulation workloads. Existing frameworks, such as Ray, are not managing this orchestration efficiently, especially in RL tasks that demand intensive input/output and synchronization between actors on a single node. In this study, we have proposed a solution implementing the reactor model, which enforces a set of actors to have a fixed communication pattern. This allows the scheduler to eliminate work needed for synchronization, such as acquiring and releasing locks for each actor or sending and processing coordination-related messages. Our framework, Lingua Franca (LF), a coordination language based on the reactor model, also supports true parallelism in Python and provides a unified interface that allows users to automatically generate dataflow graphs for RL tasks. In comparison to Ray on a single-node multi-core compute platform, LF achieves 1.21x and 11.62x higher simulation throughput in OpenAI Gym and Atari environments, reduces the average training time of synchronized parallel Q-learning by 31.2\%, and accelerates multi-agent RL inference by 5.12x.
\end{abstract}





\maketitle

\section{Introduction}
The field of machine learning (ML) has witnessed an exponential increase in computational requirements for training models, which tend to be increasingly larger deep neural networks. This complexity has necessitated the creation of new frameworks focused on training these networks and leveraging specialized hardware to reduce training times. Examples include TensorFlow~\cite{abadi2016tensorflow}, MXNet~\cite{chen2015mxnet}, PyTorch~\cite{paszke2019pytorch}, and Deepspeed~\cite{rasley2020deepspeed}. Beyond classical supervised learning, emerging AI applications are increasingly required to operate in dynamic environments and pursue long-term goals, problems that reinforcement learning (RL) is well suited for. RL is a paradigm where agents learn to make decisions by performing actions in an environment and receiving feedback in the form of rewards. This approach has already led to significant achievements such as AlphaGo~\cite{silver2016mastering}, and more recently, the success of ChatGPT~\cite{OpenAI_GPT4_2023}. RL applications span various domains, including traffic systems~\citep{chen2018auto}, UAVs~\citep{kaufmann2023champion}, large language models (LLMs)~\cite{OpenAI_GPT4_2023}, and dexterous manipulation~\citep{akkaya2019solving}.

Deep RL, unlike its traditional counterparts, is typically applied in continuous state space environments, increasing the complexity of the task and thus the computational burden. In fact, with many system optimization problems, the CPU is heavily utilized by deep RL training workloads~\cite{ali2019view}. However, the scalability of deep RL, particularly in learning complex state-action associations, hinges on efficiently leveraging both CPUs and GPUs~\cite{inci2020architectural, espeholt2019seed}. The combination of RL with deep neural networks necessitates a balanced approach in computational resource allocation. The processing speed, especially when updating policies involving millions of parameters, becomes a critical factor. The need to balance CPU and GPU resources, coupled with the limitations of existing frameworks, highlights the need for innovative approaches in developing efficient, scalable, and versatile systems capable of supporting the dynamic and complex nature of modern artificial intelligence and machine learning applications. This paper addresses these challenges and explores potential solutions, paving the way for more efficient and effective reinforcement learning systems.




To this end, we introduced an optimized version of Lingua Franca (LF)~\cite{lohstroh2021toward}, a polyglot coordination language for concurrent and time-sensitive applications. Our optimized LF is tailored to address the unique challenges of RL applications. LF stands out in its ability to effectively handle a diverse range of workloads, from lightweight, stateless computations needed for simulation to the more intensive, long-running computations required for training. A key feature of LF is its unified interface, which is adept at representing RL tasks as dataflow graphs. This offers a visual representation of the underlying RL processes, displaying the dependencies between reactors, ports, state variables, and their data. This enhances the understanding of the system's structure, which is crucial for efficiently managing the diverse computational demands of RL applications. LF also seamlessly integrates with RL workloads, including training, simulation, and serving. On a single-node multi-core compute platform, in comparison to Ray~\cite{moritz2018ray}, LF achieves 1.21x and 11.62x higher simulation throughput in OpenAI Gym~\cite{brockman2016openai} and Atari environments, reduces the average training time of synchronized parallel Q-learning by 31.2\%, and accelerates multi-agent RL inference by 5.12x.

\textbf{Contributions:} In this paper, we demonstrate how the reactor-oriented paradigm implemented in LF optimizes the utilization of computational resources and efficiently parallelizes RL workloads. We then describe the compilation process and the mechanisms of LF runtime, illustrating how reactors exploit parallelism and differ from the actors that underpin Ray. Our investigation also reveals that multithreading offers greater advantages than multiprocessing in parallel RL on a single node. To leverage these advantages, we introduced an optimized Reactor C runtime that supports Python without the Global Interpreter Lock, enabling true parallelism and abstracting away the burden of coordinating worker threads for RL tasks. We further present an extensive evaluation on the generations of samples from Open AI Gym and Atari Environments, synchronized parallel Q-learning, and inference of multi-agent RL. Our results demonstrate that LF outperforms Ray in terms of training, inference, and simulation for RL.

\section{Parallel Reinforcement Learning and Requirements}
Reinforcement Learning algorithms aim to enhance an agent's policy performance within a given environment, often represented through a simulator. These algorithms alternate cycles of data collection with the latest policy, value estimation using the latest data (and possibly data collected from previous policies), and policy improvement. Depending on the signal provided by the reward function on the collected data and the efficiency of the algorithm itself, the reinforcement learning process can vary in the amount of data and computational resources needed to run until satisfactory performance is achieved. For most tasks of interest, these data and compute requirements are large, motivating parallel reinforcement learning frameworks that can efficiently leverage multiple computational resources.

Many aspects of the reinforcement learning problem can be parallelized~\citep{samsami2020distributed}. Data collection can be split across multiple worker threads, learning can be split across multiple worker threads with each thread maintaining its own value function and policy parameters, updates to the neural networks can be performed in parallelized fashions using popular deep learning frameworks~\citep{abadi2016tensorflow, bradbury2018jax, paszke2017automatic}, and simulators can themselves be parallelized, for example on accelerated computing infrastructure~\citep{puig2023habitat3, szot2021habitat, habitat19iccv, shacklett2023extensible}. 

\subsection{Variants in RL Algorithms}

\textbf{Single-Agent Training:} The most fundamental scenario in RL is training a single agent, which involves repeatedly applying the steps of rollout, replay, and optimization. Synchronous algorithms like A2C~\citep{mnih2016asynchronous} and PPO~\citep{schulman2017proximal} follow these steps in sequence, whereas asynchronous variants (e.g., A3C~\citep{mnih2016asynchronous}, Ape-X~\citep{horgan2018distributed}, APPO~\citep{zeng2020visual}, IMPALA~\citep{espeholt2018impala}) overlap rollout and optimization steps to enhance data throughput. 

\textbf{Multi-Agent Training:} Multi-agent training involves multiple agents interacting within the environment, either cooperatively or competitively. While the dataflow structure resembles that of single-agent training, complexities arise when customizing training for individual agents. For instance, if agents require optimization at different frequencies or are trained with distinct algorithms, the training dataflow must accommodate multiple iterative loops with varying parameters. Additionally, training populations of agents, such as Double Q-learning~\cite{van2016deep}, has demonstrated great promise in RL for stabilizing training, improving exploration and asymptotic performance, and generating a diverse set of solutions.

\textbf{Model-Based Algorithms:} Model-based RL algorithms aim to learn the transition dynamics of the environment to increase training efficiency~\citep{moerland2023model}. This adds a supervised training component to parallel RL, involving training one or more dynamics models with environment-generated data. 

\subsection{Challenges and Opportunities in Framework Optimization:} 

Deep RL algorithms like Q-learning and SARSA (State-Action-Reward-State-Action) traditionally rely on sequential learning of a value table. The advent of deep neural networks has enabled these algorithms to approximate complex functions without an explicit value table. The combination of RL with deep neural networks in deep RL necessitates a balanced approach in computational resource allocation. The processing speed, especially when updating policies involving millions of parameters, becomes a critical factor. Current popular algorithms, such as Deep Q Networks (DQN) and Advantage Actor Critic (A2C), often employ a hybrid approach using both CPUs and GPUs. This setup allows for the efficient execution of different phases of the RL process, with policy evaluation typically occurring on the CPU and policy updates on the GPU. The evolution of machine learning, especially in the realm of RL, has ushered in a new era of computational requirements and challenges. 

Frameworks like Ray have made strides in CPU/GPU orchestration for RL, but they face efficiency challenges due to the actor model's inherent communication overhead and synchronization demands. Ideal frameworks for RL should support heterogeneous computations, flexibility in computational models, dynamic execution, and large data handling, while integrating seamlessly with deep learning libraries and simulation frameworks.

We argue that parallel RL frameworks should better handle millions of tasks per second by providing deterministic concurrency. Current frameworks, on the other hand, fall short in efficiently meeting the evolving demands of AI applications, indicating a significant gap in the field. This research aims to propose a novel system architecture that addresses these comprehensive requirements, bridging the gap in the current landscape and pushing the boundaries of what is possible in parallel RL.

\section{Introduction to Actor and Reactor Model}
\subsection{Actor Model}
The actor model~\cite{hewitt2010actor}, a foundational concept in concurrent computing, emerged in the 1970s as a response to the increasing complexity and interactivity in computer systems. It introduced a novel way to handle concurrent operations by conceptualizing "actors" as the primary units of computation. These actors are analogous to objects in object-oriented programming (OOP)~\cite{rentsch1982object}, but they are designed specifically to address the challenges of concurrency. Each actor represents a self-contained unit with its own local state, and actors interact with each other exclusively through asynchronous message passing. This model contrasts with traditional approaches that often rely on shared state and synchronization mechanisms like locks, which can lead to issues like deadlocks and race conditions.

Actors in the model operate independently and concurrently, providing a natural framework for distributed and parallel systems. When an actor receives a message, it can perform several actions: it can create more actors, send messages to other actors, modify its own internal state, or decide how to respond to the next message it receives. This makes the model highly adaptable and scalable, suitable for applications ranging from simple concurrent programs to complex distributed systems. The asynchronous nature of message passing in the actor model is key to its effectiveness in dealing with concurrency. It allows actors to send and receive messages without waiting for a response, thereby preventing bottlenecks and enabling continuous operation even when certain components are busy or delayed.

Furthermore, the actor model introduces a flexible approach to message processing, without enforcing any strict order in which messages must be processed. This characteristic is particularly beneficial in systems where message delivery times can vary unpredictably. Actors can process incoming messages in different sequences, and the model does not guarantee that messages will arrive in the order they were sent. This flexibility allows for more efficient utilization of resources and can lead to more robust system designs that are tolerant of delays and variable message delivery times. Overall, the actor model's emphasis on independent, concurrent actors and asynchronous communication makes it a powerful paradigm for building scalable and resilient distributed or concurrent systems.

\subsection{Reactor Model}
The reactor model~\cite{lohstroh2020reactors} shows great potential as an alternative to the actor model, enabling efficient and deterministic concurrency. The newly proposed reactor model represents an advancement in deterministic reactive systems, providing a structured framework for creating complex, reactive RL applications. Central to this model are the concepts of "reactors" and "reactions." Reactors can be interpreted as deterministic actors, but instead of responding to messages, they react to discrete events, each linked to a specific logical time, denoted by a "tag." These events can trigger reactions within a reactor, similar to message handlers in actor systems, but with a key difference: reactions in reactors are governed by a defined order, ensuring determinism. Reactions are activated by discrete events, which can also be generated by reactions. Each event associates a value with a tag, representing its logical release time within the system. Reactions can access and modify state shared with other reactions in the same reactor, but interaction between different reactors is solely through events. This design choice ensures the model's deterministic nature because the order of reaction executions is predictable and subject to strict constraints, such as tag order and execution order for reactions within the same reactor.

The reactor model can be thought of as a "sparse synchronous model" \cite{EdwardsHui:20:SSM}. This means that synchronous-reactive interactions at a particular logical time may be confined to isolated parts of the system. When a reaction executes, it exclusively accesses the reactor's state. Moreover, for reactions in the same reactor that are triggered at the same tag, their execution order is predefined, enforcing deterministic behavior. Reactors are composed of ports (inputs and outputs), hierarchy, local state, and actions. The term "reactors" not only relate to actors in actor systems but also aligns with the synchronous reactive programming paradigm, prominent in languages like Esterel, Signal, and Lustre \cite{Benveniste:91:Synchronous}. Unlike traditional actors, reactors don't directly reference their peers. Instead, they use named and typed ports for interconnection, enabling a hierarchical design. This hierarchy, besides facilitating deterministic behavior, also serves as a scoping mechanism for ports and imposes constraints on connection types.

Additionally, reactors feature actions, a variant of ports used for scheduling future events within the same reactor or as a synchronization mechanism between internal logic and asynchronous external events. This design choice provides a bridge between deterministic internal logic and the nondeterministic external world, like sensor data, environment states, or network messages. The reactor model also incorporates state variables. The shared resources, like replay buffers, model parameters, environment states in RL, are key motivators for grouping reactions in a single reactor. However, reactors themselves do not share state, ensuring isolation between reactors and allowing parallel execution of reactions in different reactors, unless a connection necessitates sequential execution.

Connections between reactors establish explicit communication channels. These connections reveal dependencies that are crucial for scheduling decisions honoring data dependencies. The reactor model simplifies the declaration of these dependencies by breaking down functionality into reactions with well-defined lexical scopes, thus eliminating dependencies out of scope. The reactor model's execution is governed by a run-time environment. This environment is responsible for maintaining a global event queue and a reaction queue, managing logical time, and executing reactions. Details regarding the scheduler are described below.

\subsection{Ray}
Ray~\cite{liang2021rllib, moritz2018ray} is an open-source framework designed for distributed computing and has been widely used to parallelize RL workloads from single-node to multi-nodes. Ray utilizes the actor model, which is central to its architecture and operation, making it highly effective for concurrent and distributed computing tasks. The actor model in Ray is used to encapsulate state and behavior, with actors being distributed across a node or a cluster and communicating through asynchronous message passing. It is important to highlight that in this study, our primary focus is on enhancing the efficiency of distributing actors within a node. Actors can be used for performance reasons (like caching soft state or ML models), or they could be used for managing long-living connections to databases or to web sockets. They can maintain state across multiple tasks, which is particularly useful for applications that require managing large, mutable states, such as machine learning models or large datasets.

The local scheduler in Ray is a component of a worker node called Raylet. Raylet manages the worker processes and consists of two components, a task scheduler and an object store. The task scheduler takes care of scheduling and executing work on a node. It addresses issues such as a worker being busy, not having the proper resources to run a task, or not having the values it needs to run a given task. Expensive serialization and deserialization as well as data copying are common performance bottleneck. Shared memory, specifically that which is managed directly by the operating system kernel, emerges as a superior mechanism compared to conventional approaches like socket connections. The fundamental attribute of the in-memory object store is that all objects within the store are immutable and retained in shared memory. The object store has its eviction policy, removing objects from the store or transferring them to other nodes when the allocated size limit is exceeded. This design choice ensures optimal access speeds, particularly when multiple workers on a singular node need to engage with the data. Each node provisioned by Ray is equipped with an object store, within that node’s Raylet. Functionally, the object store takes care of memory management and ultimately makes sure workers have access to the objects they need in Ray.

\subsection{Lingua Franca}
Lingua Franca (LF)~\cite{menard2023high, schulz2023polyglot} is a polyglot coordination language designed to facilitate the development of concurrent systems by focusing on deterministic interactions with the environments. It operates on the reactor model, where reactors are the fundamental units of composition, each encapsulating reactions to external stimuli. LF's main feature is its deterministic nature, meaning that given a set of inputs, a LF program will always produce the same outputs, greatly enhancing testability and efficiency. This determinism is achieved by using a superdense model of time (where events are ordered by time and microstep), ensuring causality and the absence of non-deterministic feedback loops in the reaction network.

The architecture of LF includes a compiler (lfc). The compiler process involves parsing and validating the LF code, checking for syntax errors, instantiation cycles, and cyclic dependencies among reactions. Valid code can then be transpiled into target code in languages like C, C++, TypeScript, and Python, which is then combined with a runtime system to manage the execution of reactors. LF also allows for graphical program representation, enhancing program structure understanding and error identification.

In LF, reactors can be defined with parameters (immutable after initialization), ports (for data input/output), actions (for scheduling internal events), and timers (for periodic events). Reactors can also declare state variables to maintain state across logical time. Reactions in LF, which contain the core logic, are defined with triggers (conditions under which they execute), sources (additional data inputs), and effects (outputs or actions they can trigger). LF supports reaction deadlines, where an alternative code block executes if a reaction misses its specified deadline, thus enforcing timing constraints.

LF allows for flexibility in connecting reactors through multiports (ports handling multiple data channels) and banks (multiple reactor instances). Connections can be logical, implying synchronization between ports, or physical, introducing intentional nondeterminism for scenarios where strict event ordering is not required. LF's syntax permits the creation of complex interaction patterns among reactors, offering a robust toolset for constructing scalable and maintainable concurrent systems.

The formal semantics of LF is grounded in the theory of discrete-event systems, utilizing a generalized ultrametric space for modeling the behavior of LF programs~\cite{Liu:06:Metric}. This approach guarantees that each LF program is deterministic and adheres to causality, essential for reliability in concurrent system design. LF's semantic model is fully abstract, providing both an operational perspective (how the program executes) and a denotational perspective (the meaning of program constructs), ensuring consistency. An operational semantics of the reactor model based on a formalization in LEAN~\cite{de2015lean} is also available~\cite{Rossel2023}.

\begin{figure}
    \centering
    \includegraphics[width=1.0\linewidth]{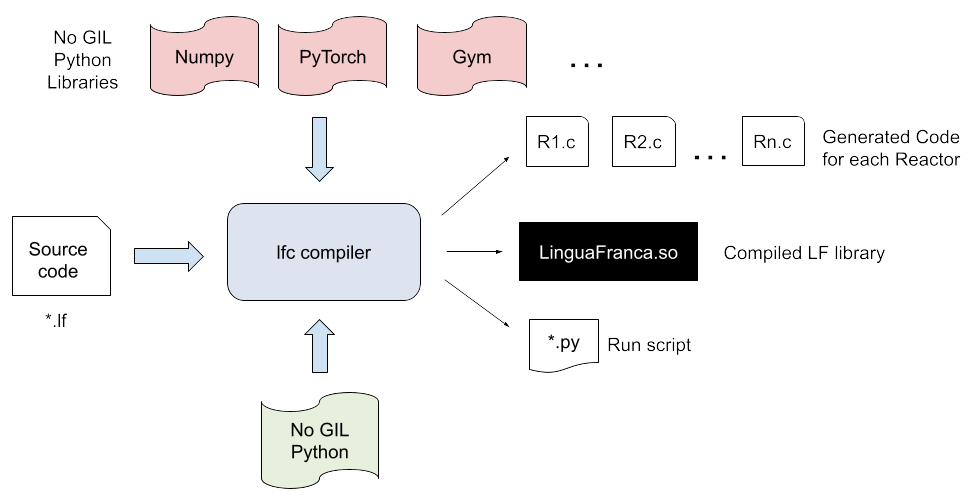}
    \caption{LF: compilation process}
    \label{fig:LF}
\end{figure}

Fig. ~\ref{fig:LF} depicts the compilation process of the LF framework. This process begins with the source code written in Lingua Franca, with a .lf extension. The source code is then processed by a modified lfc compiler, tailored to work without Python's Global Interpreter Lock (GIL). Additionally, the compilation process involves various no-GIL versions of the Python libraries such as NumPy, PyTorch, Gym, etc. 

This compiler generates C code for each reactor, with files named R1.c, R2.c, and so on as well as the compiled Lingua Franca library, LinguaFranca.so, which can be imported and used in Python scripts (*.py). The runtime implementation serves as a bridge between the high-level coordination language and the underlying Python environment without the GIL, enabling users to write truly concurrent Python programs. It abstracts away the burden of coordinating worker threads.

\subsection{Representing RL tasks as a Dataflow Graph}

\begin{figure}
    \centering
    \includegraphics[width=0.9\linewidth]{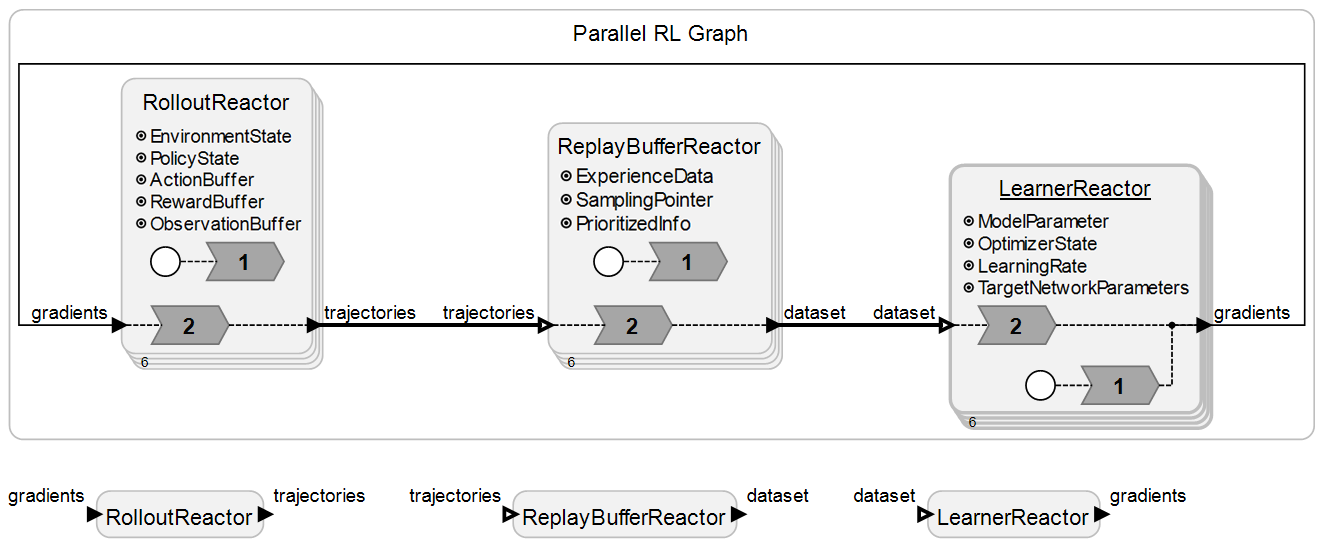}
    \caption{Generated Dataflow Graph for parallel RL tasks}
    \label{fig:dataflow}
\end{figure}

The diagram synthesizer in LF provides a streamlined process to represent parallel RL tasks as dataflow graphs~\cite{lohstroh2021toward}. By simply initializing reactors and setting their input and output ports, LF can automatically generate the corresponding dataflow graph. This diagrammatic feature is seamlessly integrated with Visual Studio Code and Eclipse, offering a visual representation of the underlying RL processes. The resulting diagram (Fig.~\ref{fig:dataflow}) clearly displays the reactors, ports, state variables, and their data dependencies, simplifying the understanding of the system's structure and aiding in the debugging of parallel RL tasks. Users are therefore better positioned to manage the parallelization of RL processes. LF also offers a compact syntax for ports that are capable of sending or receiving across various channels, and a syntax for multiple instances of a reactor. These concepts are known as multiports and banks of reactors. For example, we have created a bank of six instances of ReplayBufferReactor and one instance of LearnerReactor, connecting them using multiport. Detailed information about the implementation is provided in the appendix. 

\textbf{RolloutReactor}: This reactor is pivotal in interacting with the environment. It gathers trajectories by executing a policy and recording the resulting states, rewards, and other pertinent outcomes. It receives gradients from the LearnerReactor for policy updates and sends experience (e.g. trajectories) to the ReplayBufferReactor.

\begin{itemize}
    \item \textit{EnvironmentState}: The present state of the environment.
    \item \textit{PolicyState}: The weight of the current policy.
    \item \textit{ActionBuffer}: A temporary repository for actions decided by the policy.
    \item \textit{RewardBuffer}: Temporary storage for rewards post-action.
    \item \textit{ObservationBuffer}: Temporary storage for new environmental observations post-action.
\end{itemize}

\textbf{ReplayBufferReactor}: Acting as a centralized experience replay buffer, this reactor stores trajectories from the RolloutReactor for subsequent sampling. It provides batched experience to the LearnerReactor for policy updates. 

\begin{itemize}
    \item \textit{ExperienceData}: An accumulation of experiences (state, action, reward, subsequent state, and termination info).
    \item \textit{SamplingPointer}: Indices or pointers facilitating efficient sampling.
    \item \textit{PrioritizedInfo}: For prioritized replay buffers, additional information will be used to manage sampling.
\end{itemize}

\textbf{LearnerReactor}: The LearnerReactor updates policies based on sampled experiences from the ReplayBufferReactor or directly from the RolloutReactor, and broadcasts updated gradients to the RolloutReactor. 

\begin{itemize}
    \item \textit{ModelParameter}: The neural network parameters, including weights and biases.
    \item \textit{OptimizerState}: Elements related to the optimization process, such as momentum variables and correction terms.
    \item \textit{LearningRate}: The current learning rate, either static or dynamically adjusted.
\end{itemize}
\begin{itemize}
    \item \textit{TargetNetworkParameters}: In algorithms like DQN~\cite{mnih2013playing}, these are slowly updated parameters offering a stable learning target.
\end{itemize}
 
\section{Optimizations for Parallel Reinforcement Learning}

\subsection{Scheduling Algorithm}
\begin{figure}
    \centering
    \includegraphics[width=0.9\linewidth]{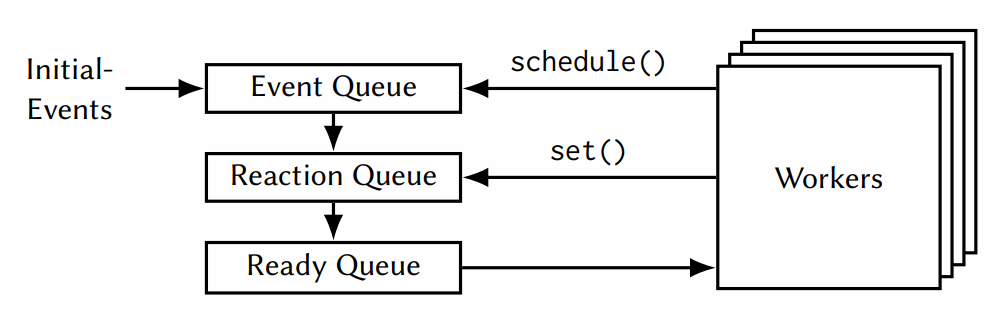}
    \caption{Scheduling mechanism in the LF runtime ~\cite{menard2023high}}
    \label{fig:schedule}
\end{figure}

The execution process of programs in LF involves a scheduler responsible for overseeing all scheduled future events, managing the logical time progression, and executing triggered reactions in the order dictated by the dependency graph. The scheduling mechanism in LF, depicted in Fig.~\ref{fig:schedule}, operates with an event queue that strictly follows a tag order for processing upcoming events. Upon adding an event to the queue and processing it, the scheduler identifies triggered reactions, placing them in a reaction queue. These reactions are subsequently moved to a ready queue and executed by worker threads once all dependencies, as outlined by the Action-Port Graph (APG), are satisfied. Within this scheduling framework, the scheduler ensures a sequential execution of reactions that depend on one another, aiming to maximize parallelism.

The scheduling mechanism is similar to Directed Acyclic Graph (DAG)-based strategies but differs by accommodating reactions within the Action-Port Graph (APG) that may not always need execution. Since the scheduler cannot predict in advance which reactions will be triggered at a given tag, it cannot precompute an optimal schedule. To address this, the scheduler assigns a level to each reaction, allowing reactions at the same level to execute concurrently. This heuristic eliminates the need for runtime APG analysis, streamlining the process. The scheduler processes reactions sequentially, advancing to the next level only after completing all reactions at the current level. This approach substantially reduces synchronization overhead and contention on shared resources, contributing to its efficiency.

\subsection{Scheduling Optimizations}
The scheduling algorithm described in the above section is relatively straightforward to implement, but achieving competitive performance requires additional optimizations. These optimizations, implemented in the C runtime in LF, are detailed below.

\textbf{Coordination of worker threads}: Conceptually, the scheduler and workers are distinct; however, in practice, having a central scheduler with separate worker threads can cause significant synchronization overhead. To mitigate this, our implementation allows any worker thread to become the scheduler, allowing it to move ready reactions to the queue or advance to the logical time once all reactions are processed. Additionally, LF leverages the fact that we know the number of parallel reactions to execute from the APG, and thus can use a counting semaphore to regulate the number of active workers.

\textbf{Lock-free data structure}: The three queues (event, reaction, and ready queue) and other data structures are shared across all workers. Mutex-based synchronization would be inefficient due to high contention over these shared resources. Accordingly, LF adopts lock-free data structures wherever feasible. For example, the ready queue is a fixed-size buffer with an atomic counter, which corresponds to the maximum number of parallel reactions defined by the APG level. Each time a worker tries to execute a reaction it atomically decrements the counter. If the counter is negative, it indicates an empty queue, and the worker should proceed accordingly. Otherwise, the counter provides the index within the buffer from which to read. This operation is safe without additional synchronization, as all workers are waiting for upcoming new reactions.

\subsection{Enabling Multi-threading for Reinforcement Learning}
Parallel processing is a cornerstone in the field of machine learning, enabling the handling of computationally intensive tasks and large-scale data. The parallel execution models, namely multithreading and multiprocessing, provide different advantages. This section aims to dissect these models to guide the selection of an appropriate parallel execution strategy for RL.

\textbf{Multithreading in RL:} 
\begin{enumerate}
    \item Shared Memory Space: Multithreading involves threads operating within the same memory space, which facilitates faster and more efficient data sharing among threads compared to multiprocessing.
    \item Resource Efficiency: The creation and management of threads consume fewer resources than processes. This efficiency stems from the shared memory space and the absence of a need for complex inter-process communication mechanisms.
    \item I/O Bound Task Optimization: Multithreading proves advantageous for I/O-bound operations, where the ability to perform other tasks while waiting for I/O operations enhances efficiency.
    \item Context Switching: The shared process and memory space of threads enable faster context switching than multiprocessing, as less information needs to be saved and restored.
\end{enumerate}

\textbf{Multiprocessing in RL}
\begin{enumerate}
    \item CPU Bound Task Optimization: Multiprocessing is typically more suitable for CPU-bound tasks, allowing for the distribution of tasks across multiple CPU cores. However, most of the CPU-bound tasks (e.g. gradient updates) in RL should be offloaded to GPU instead.
    \item Fault Tolerance and Stability: The isolated nature of processes in multiprocessing ensures that a crash in one process does not impact others, thus enhancing application stability. The scheduler in LF guarantees the safety of threads and permits users to specify actions in the event of a failure.
    \item Bypassing the GIL in Python: In Python, the Global Interpreter Lock (GIL) limits thread execution, multiprocessing provides a viable alternative for parallel CPU computations. This has been resolved by leveraging the No GIL version of Python. 
    \item Multi-Core Utilization: Multiprocessing enables efficient utilization of multi-core processors by running separate processes on each core. However, it's important to consider that when sending excessively large objects, the processor may spend a significant amount of time on I/O overhead. 
\end{enumerate}

\subsection{Optimizing Thread Allocations}
The principle that running a task with multiple threads can speed up the process is generally true, as each thread can handle a portion of the work simultaneously. However, the scaling is not always linear due to various factors like thread allocation and CPU architecture. In our benchmarks, the Intel CPU has a hybrid architecture, featuring a combination of performance and efficiency cores. Performance cores are designed for high-speed and intensive tasks, while efficiency cores are optimized for lower power consumption and handling background tasks. The original LF runtime randomly allocates threads for workers. As a result, when synchronous tasks are distributed across these cores, the overall speed is bounded by the slower efficiency cores. To achieve linear scaling, the Reactor C runtime has been optimized to prioritize using the performance cores before spawning threads. 

\begin{figure}[ht]
    \centering
    \includegraphics[width=0.65\linewidth]{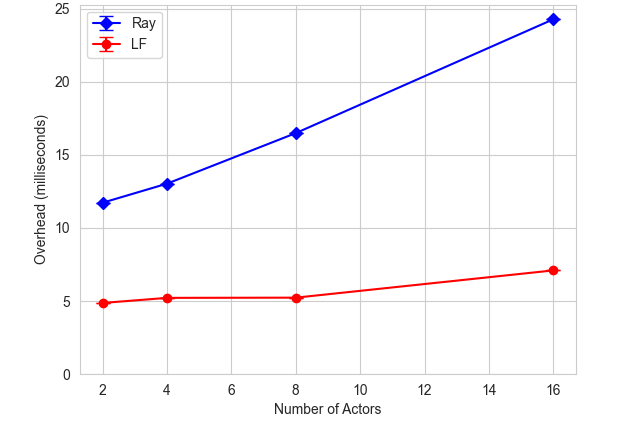}
    \caption{Mean Overhead of Broadcast and Gather 10MB Object with Different Number of Actors using Ray and LF.}
    \label{fig:num-actors}
\end{figure}

\begin{figure}[ht]
    \centering
    \includegraphics[width=0.65\linewidth]{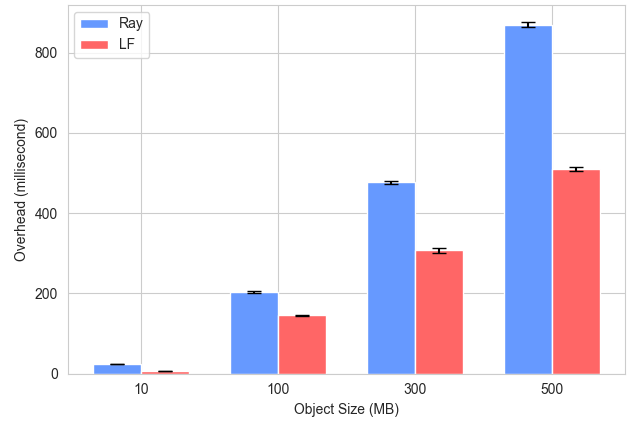}
    \caption{Mean Overhead of Broadcast and Gather on 16 actors with Different Object Sizes using Ray and LF.}
    \label{fig:object-size}
\end{figure}

\section{Performance Comparison Between LF and Ray}

To demonstrate the improved performance of LF over Ray for parallel RL workloads, we run the following experiments. We believe these experiments are comprehensive and adequate to support our assertion that our optimized implementation of the reactor model for RL significantly advances the state of the art for parallel RL.
\begin{enumerate}
    \item We demonstrate lower overhead of broadcast and gather operations with LF than with Ray, and show that the overhead difference scales favorably for LF as we increase both the number of actors/reactors and the communication object size. While not an RL benchmark, this toy workload resembles the computation done in RL, and is useful for highlighting why exactly it is that LF outperforms Ray (sections \ref{exp:num-actors}, \ref{exp:obj-size}).
    \item In most parallel RL settings, (e.g., as is typically the case with policy gradient and Q-learning algorithms), the majority of the parallelism that can be obtained by porting these algorithms to leverage parallel compute is in the data collection phase of the algorithm (as opposed to for example value estimation or policy improvement). As such, we extensively test parallel data collection in popular RL simulated environments with both LF and Ray (sections \ref{exp:gym_envs}, \ref{exp:atari_envs}).
    \item To demonstrate that LF integrates seamlessly with other forms of parallel compute paradigms, namely GPU acceleration, and to demonstrate the versatility LF provides to implement various reinforcement learning algorithms, we implement and evaluate synchronized parallel Q-learning with deep neural networks (section \ref{exp:dqn}).
    \item Finally, we observe that multi-agent RL lends itself very well to being parallelized, as each agent can maintain its own learning and data collection actors/reactors. We evaluate in such a multi-agent RL (MARL) setting and observe favorable results for LF (section \ref{exp:marl}).
\end{enumerate}

\subsection{Experimental Setup}
The actor model is a popular approach for developing parallel RL applications, with Ray being notable for its usability and efficiency in handling multiple actors. However, LF introduces a different model of computation, imposing more restrictions compared to the actor model. We hypothesize that the reactor model with LF can outperform Ray, which is very widely used today for parallel RL. Drawing evidence from this study~\cite{menard2023high}, the reactor model has been shown to surpass the performance of the traditional Actor Framework, Akka and C++ Actor Framework, by factors of 1.86x and 1.42x, respectively, for non-RL workloads. More specifically, the reactor model with a fixed set of actors and fixed communication patterns allows the scheduler to eliminate work needed for synchronization. Furthermore, with our customized implementation, LF leverages No GIL Python for multithreading, which offers several benefits, including efficient data sharing within a shared memory space and faster context switching. The reduced overhead is demonstrated by both Fig.~\ref{fig:num-actors} and Fig.~\ref{fig:object-size} . The term, overhead, refers to the duration required for one actor to send a payload to another and for it to be received. This process may include serialization and deserialization, acquiring and releasing locks for each actor, sending and processing coordination-related messages, as well as transferring data over the network. All measurements were performed on AWS EC2 m5.8xlarge instance, equipped with an Intel® Xeon® Platinum 8175M CPU @ 2.50GHz featuring 32 vCPUs. This setup includes 128 GiB of RAM and offers a 10 Gbps network bandwidth. The system runs on Ubuntu 20.04 and use Python 3.9.10, with NumPy version 1.22.3 and gym version 0.19.0.

\subsection{Number of Actors}
\label{exp:num-actors}
Fig.~\ref{fig:num-actors} illustrates the mean overhead of broadcasting and gathering a 10MB object across different numbers of actors in a parameter server setup. Two frameworks are compared: Ray and LF. The x-axis represents the number of actors, which are 2, 4, 8, and 16, while the y-axis shows the overhead in milliseconds. Both frameworks exhibit an increase in overhead as the number of actors grows, but Ray consistently has a higher overhead than LF. For instance, with 16 actors, Ray's overhead is close to 20 milliseconds, whereas LF's is just above 5 milliseconds.

\begin{figure}[ht]
    \centering
    \includegraphics[width=0.6\linewidth]{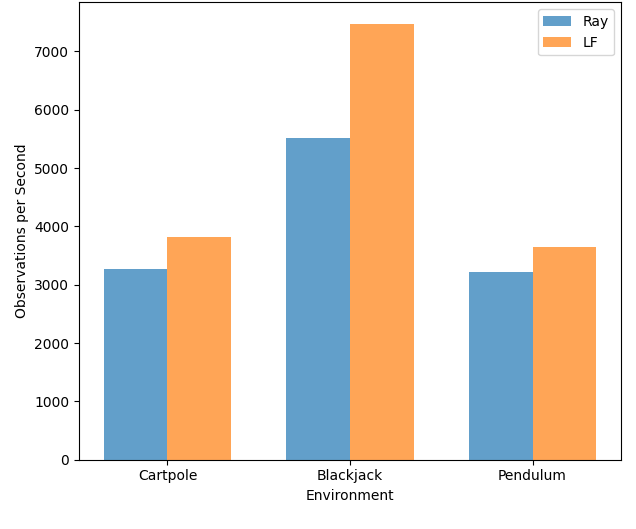}
    \caption{Simulation Throughput of Ray and LF with 16 actors in Various Gym Environments.}
    \label{fig:sim-gym}
\end{figure}

\begin{figure}[ht]
    \centering
    \includegraphics[width=0.6\linewidth]{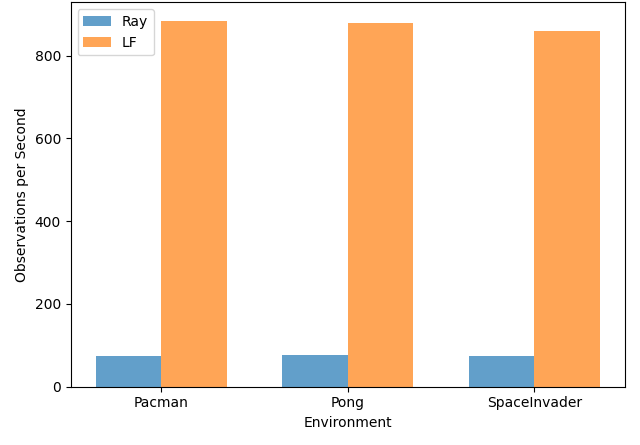}
    \caption{Simulation Throughput of Ray and LF with 16 actors in Various Atari Environments.}
    \label{fig:sim-atari}
\end{figure}

\subsection{Object Size}
\label{exp:obj-size}
Fig.~\ref{fig:object-size} presents a comparison of the mean overhead for broadcasting and gathering operations on objects of varying sizes using 16 actors, between Ray and LF frameworks, including a 99\% confidence interval (CI). The x-axis displays the object size in megabytes (MB), ranging from 0 to 500 MB, while the y-axis indicates the overhead in milliseconds. From the graph, we can observe that as the object size increases, the overhead for both Ray and LF also increases. However, Ray's overhead grows at a higher rate than LF's. For instance, with the largest object size of 500 MB, Ray's overhead approaches 800 milliseconds, whereas LF's overhead is about half of that, around 400 milliseconds.

Both Fig.~\ref{fig:num-actors} and Fig.\ref{fig:object-size} support the hypothesis that the reactor model employed by LF can outperform traditional actor models like Ray, particularly in scenarios with a greater number of actors and larger object size, due to optimizations such as reduced synchronization work and efficient multithreading with No GIL Python.

\subsection{Open AI Gym Environments}
\label{exp:gym_envs}
In Fig.~\ref{fig:sim-gym}, we see that LF outperforms Ray in terms of simulation throughput for Open AI Gym Environments by 1.21x on average, with a particularly significant lead in the Blackjack environment. This suggests that LF is more efficient, especially in situations where there is a lower CPU demand for action updates within the environment. It's important to note that vectorized environments are asynchronous and do not parallelize inference of policy. Therefore, it is not included in our benchmarks.

\subsection{Atari Environments}
\label{exp:atari_envs}
In Fig.~\ref{fig:sim-atari}, we compare the performance of Ray and LF on Atari environments, specifically on Pacman, Pong, and SpaceInvader. Here, LF again significantly outperforms Ray in terms of observations per second. On average, LF is roughly 11.62x faster than Ray across these environments. The substantial performance difference can be attributed to LF's efficiency in handling high I/O (input/output) bound tasks. Atari environments are more complex than the previously mentioned OpenAI gym environments. They represent each state as an 80x80 numpy array, which requires more computational resources to serialize and deserialize, especially when data needs to be sent over a network. Ray's use of pickle5 for serialization does help to increase throughput by efficiently serializing NumPy arrays, but it still introduces overhead during network transmission and the serialization/deserialization process. This overhead is particularly significant in environments where state updates are frequent and must be communicated quickly. LF's ability to handle the demands of complex simulation environments is thus a key advantage.

\begin{figure}[ht]
    \centering
    \includegraphics[width=0.9\linewidth]{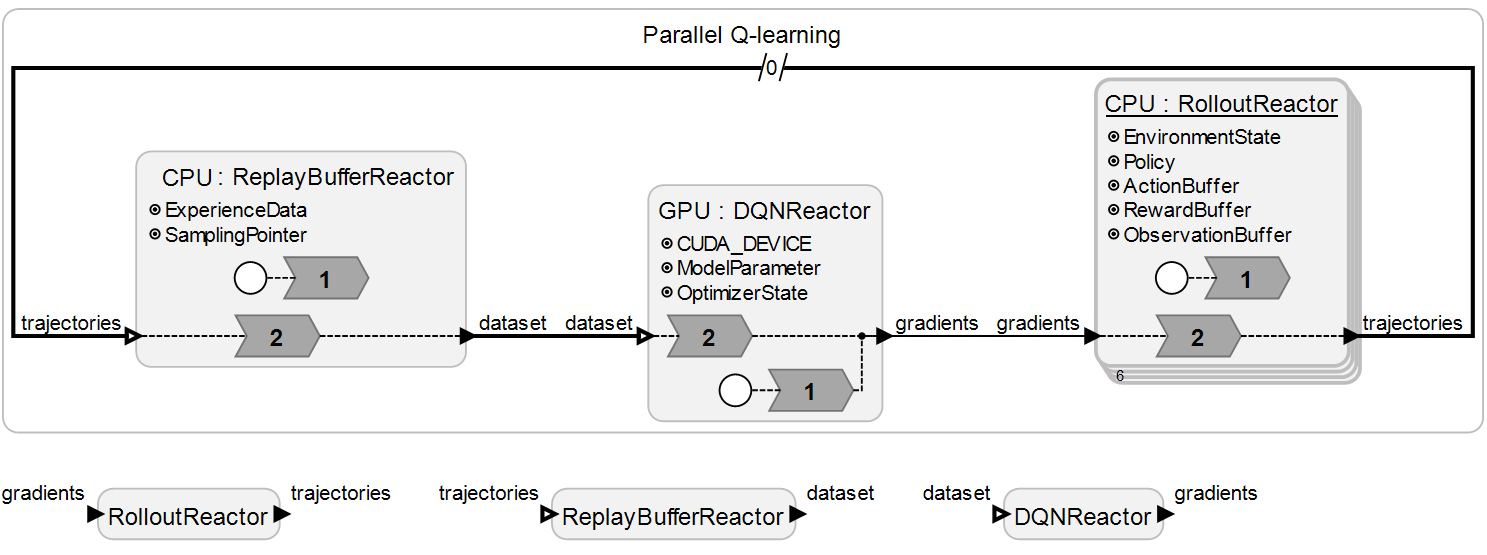}
    \caption{Dataflow Graph of Parallel Q-learning}
    \label{fig:dataflow-dqn}
\end{figure}

\subsection{Synchronized Parallel Q-learning}
\label{exp:dqn}
Deep Q-Networks (DQNs) ~\cite{mnih2013playing} are an advancement in reinforcement learning that utilize deep neural networks to estimate Q-values. The Q-values are predictions of the expected discounted returns after taking certain actions given particular states in an environment. DQNs extend the capabilities of traditional Q-learning by handling larger state and action spaces, which are common in complex problems. These networks can scale up effectively with more data or increased model complexity; therefore, DQN usually utilizes GPUs for gradient updates. However, since AWS EC2 m5.8xlarge does not include a GPU, benchmarking was conducted on a workstation with an Intel i9-13950HX CPU @ 2.20GHz featuring 32 vCPUs, NVIDIA RTX4090, and 32 GiB of RAM. The system also runs on Ubuntu 20.04 and uses Python 3.9.10, with numpy 1.22.3, torch 1.9.0, and gym 0.20.0.

DQN has been implemented in a synchronized parallel Q-learning setup within a BlackJack environment~\cite{silver2013playing}. The dataflow graph is shown in Fig.~\ref{fig:dataflow-dqn}. The network is trained to take a blackjack hand as input and output scores for each of the possible actions in the game, which represent the expected rewards of taking those actions. The computational tasks are distributed with the rollout and replay buffer being executed on a CPU, while the DQN Reactor uses a GPU. This setup takes advantage of GPUs for complex matrix operations and multicore CPUs for sequential decision-making simulations in Open AI Gym environments.

\begin{figure}[ht]
    \centering
    \includegraphics[width=0.7\linewidth]{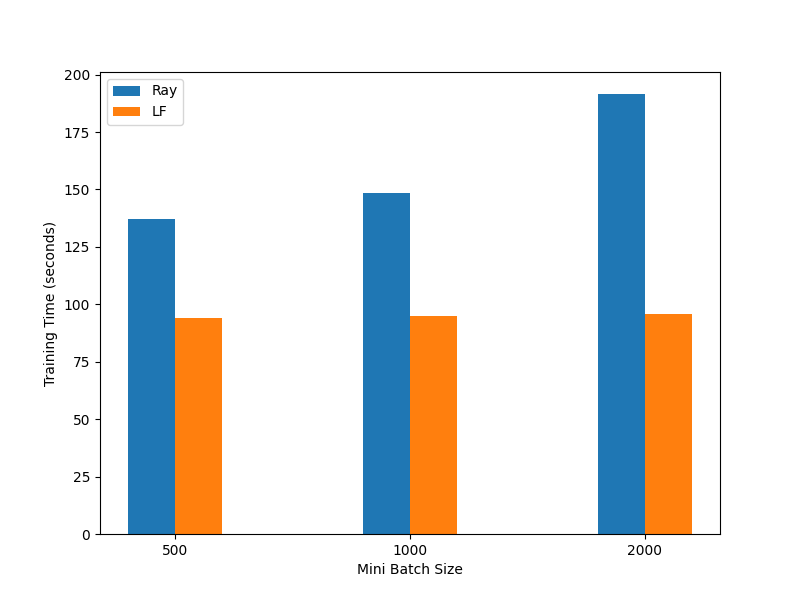}
    \caption{Synchronized Parallel Q-learning with Different Batch Sizes from a Replay Buffer}
    \label{fig:parallel-dqn}
\end{figure}

The benchmark results, as seen in Fig.~\ref{fig:parallel-dqn}, demonstrate performance improvements using LF over Ray. In tests with a 500 sample batch size from the replay buffer, the average training time decreased by 31.2\%. The data also shows that while Ray's training time increases with larger mini-batch sizes, LF's performance remains stable, indicating its ability to handle larger batches efficiently.



\subsection{Multi-Agent RL Inference Comparison}
\label{exp:marl}
In Multi-Agent RL (MARL), multiple agents operate in a common environment, where each of them tries to optimize its own return by interacting with the environment and other agents~\cite{zhang2021multi}. In centralized MARL, a central controller aggregates information across the agents, including joint actions, rewards, and observations, and policies across different agents. In decentralized MARL, which is more common in cooperative situations, each agent makes decisions based on its local observations~\cite{oroojlooy2023review}. 

\begin{figure}[ht]
    \centering
    \includegraphics[width=0.65\linewidth]{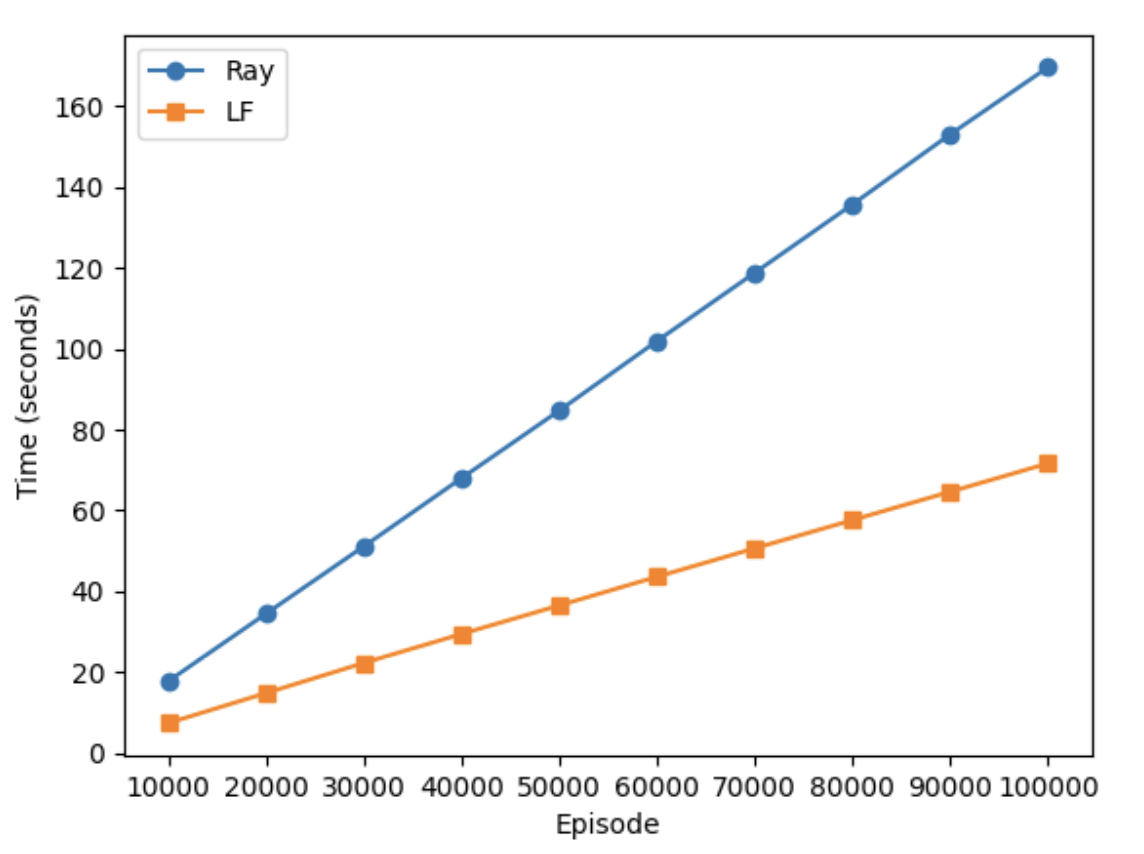}
    \caption{Inference Time Between Ray and LF over Episode Counts}
    \label{fig:episode}
\end{figure}

\begin{figure}[ht]
    \centering
    \includegraphics[width=0.65\linewidth]{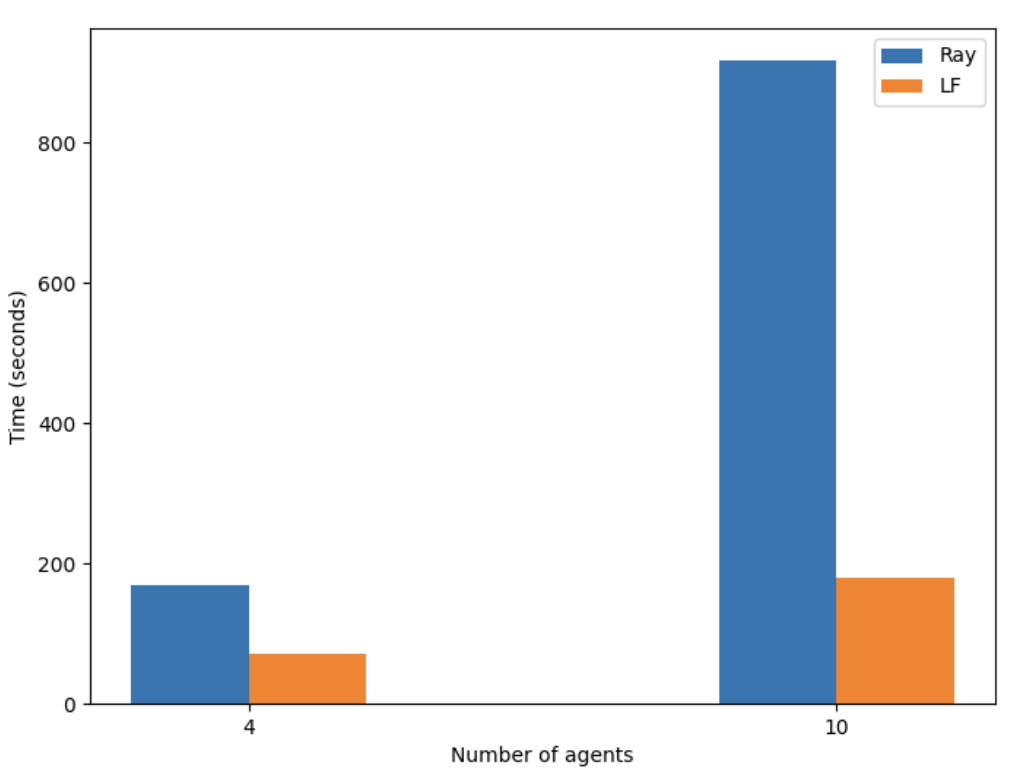}
    \caption{Inference Time Comparison between Ray and LF across Various Numbers of Agents}
    \label{fig:num_agents}
\end{figure}

We validate LF in a MARL setting with \textit{ma-gym}~\cite{magym}, a MARL library based on OpenAI Gym~\cite{brockman2016openai}. We use the TrafficJunction4-v0 environment, in which four agents are trying to pass a crossroad without crashing. This is a decentralized setting where each agent gets its own local observation. As shown in Fig.~\ref{fig:episode}, LF requires significantly less inference time than Ray (less than 1/2 the fitted line's slope). It is also noteworthy that as the number of agents increases as shown in Fig.~\ref{fig:num_agents}, LF's inference time scales better (again, less than 1/2 the rate of increase than that of Ray). In TrafficJunction environments with 10 agents, LF achieves a 5.12x speed-up compared to Ray.

\section{Discussion}
We demonstrate that LF outperforms widely used frameworks like Ray in handling training, serving, and simulation tasks in RL. We achieve this by reducing the work needed for synchronization using the reactor model and decreasing the I/O overhead through optimizing the coordination of Python worker threads. Our empirical evaluations demonstrate LF's superior performance: a 1.21x and 11.62x higher simulation throughput in OpenAI Gym and Atari environments, a 31.2\% reduction in average training time for synchronized parallel Q-learning, and a 5.12x acceleration in multi-agent RL inference. We aim to incorporate the optimizations for single-node described in this work into LF’s federated execution, enabling efficient distributed training and serving across nodes. We also plan to delve deeper into the potential applications of our optimizations in deploying deep RL on embedded systems, and will compare it with frameworks such as the Robot Operations System, a middleware commonly used in robotics.

\bibliographystyle{ACM-Reference-Format}
\bibliography{references}










\end{document}